\documentstyle[epsfig,a4,12pt]{article} 

\def\tou#1{{\lower1.2ex\hbox{$\longrightarrow$}\atop
        {\lower-.7ex\hbox{$\scriptscriptstyle #1 $}}}}
\def\lsim{{\lower1.2ex\hbox{$<$}\atop
        {\lower-.7ex\hbox{$\sim$}}}}
\def\gsim{{\lower1.2ex\hbox{$>$}\atop
        {\lower-.7ex\hbox{$\sim$}}}}

\def\be{\begin{equation}}
\def\ee{\end{equation}}

\begin{document}

\begin{titlepage}
%\rightline {Si-95-13 \  \  \  \   }

\vspace*{2.truecm}

\centerline{\Large \bf  Generalized Dynamic Scaling}
\vskip 0.6truecm
\centerline{\Large \bf 
        for Critical Relaxations }
\vskip 0.6truecm

\vskip 2.0truecm
\centerline{\bf B. Zheng}
\vskip 0.2truecm

\vskip 0.2truecm
\centerline{Universit\"at -- GH Siegen, D -- 57068 Siegen, Germany}

\vskip 2.5truecm

\abstract{ The dynamic relaxation process for the two dimensional
Potts model at criticality starting from an initial state 
with very high temperature and arbitrary magnetization is investigated
with Monte Carlo methods. The results show that 
there exists universal
scaling behaviour even in the short-time regime of the dynamic evolution.
In order to describe the dependence of the scaling
behaviour on the initial magnetization, a critical characteristic
function is introduced.

}


\end{titlepage}

For long it was believed that no universal behaviour would be present
in the short-time regime of critical dynamics. 
However, for the critical relaxation process starting from an initial state
with {\it very high temperature} and {\it small magnetization}, 
it was recently argued by Janssen, Schaub and Schmittmann \cite {jan89}
with renormalization group methods that 
there exist
universality and scaling even {\it at early times},
 which sets in right after a
microscopic time scale $t_{mic}$. 
For the $O(N)$ vector model with dynamics of
model A,  the authors derived
a dynamic scaling relation which is valid up to the macroscopic
short-time regime,
\begin{equation}
M^{(k)}(t,\tau,m_{0})=b^{-k\beta/\nu}
M^{(k)}(b^{-z}t,b^{1/\nu}\tau,
b^{x_{0}}m_{0})
\label{e5}
\end{equation}
where $M^{(k)}$ is $k$-th moment of the magnetization,
$t$ is the dynamic evolution time, $\tau$ is the reduced temperature,
and the parameter $b$ represents the spatial rescaling factor. 
Besides the well known static critical exponents $\beta$, $\nu$
and the dynamic exponent $z$, 
a new independent exponent $x_0$ which is
the anomalous dimension
of the initial magnetization $m_0$ is introduced
to describe the dependence of the scaling behaviour on
the initial conditions. 
Based on the scaling relation it is 
predicted that at the beginning of the time evolution the magnetization
 surprisingly 
undergoes a {\it critical initial increase}
\begin{equation}
M(t) \sim m_0 \, t^\theta
\label{e10}
\end{equation}
where the exponent $\theta$ is related to the exponent $x_0$ by
$\theta=(x_0-\beta/\nu)/z$.
This makes the effect of $m_0$ very prominent.

Numerical simulation supports the above predictions
for the critical short-time dynamics. 
The exponent $\theta$ for the Ising model
was first obtained indirectly
through the power law decay of the auto-correlation \cite {hus89,hum91}.
Recently the initial increase of the magnetization in 
(\ref {e10}) was observed
for the Ising model and the Potts model, and the exponent $\theta$
was directly measured \cite {li94,sch95}. The scaling relation
and the universality are confirmed \cite {li94,men94,oka95}. 
The investigation of the universal behaviour of 
the short-time dynamics not only enlarges the fundamental knowledge
on critical phenomena but also, more interestingly, 
provides possible new ways to determine all the 
static exponents as well as the dynamic exponents
from the short-time dynamics,
either based on the power law behaviour of the observables at the beginning
of the time evolution \cite {sch95,sch95a,cze95}, 
or on the finite size scaling \cite {li95,li95a}.
Moreover the universal behaviour of the short-time dynamics is found to be
quite general, e.g. in the dynamics beyond model A or 
at the tri-ciritical point \cite {jan92,oer93,oer94}, 
in connection to ordering dynamics or 
damage spreading \cite {bra94,gra95} and on the surface critical phenomena
\cite {rit95}. 
Therefore thorough understanding of the universality
and scaling for the short-time dynamics is urgent and important.

The scaling relation 
(\ref {e5}) 
is valid only under the conditions that the initial state
is at very high temperature and with {\it small} initial magnetization.
Are there some reasons that the universal behaviour
 emerge only in the critical relaxation starting from
such a special initial state? If one believes that
the large time correlation length is 
essential for the universality, 
the only background one would
find is that the initial temperature $T_0=\infty$ and the 
initial magnetization $m_0=0$ are the fixed points under
the renormalization group transformation. 
Therefore one may wonder whether there exists universal behaviour  
in the critical relaxation process starting from an initial
state with very high temperature but initial magnetization 
$m_0 \simeq 1$, since $m_0=1$ also corresponds to
a fixed point. Actually the critical relaxation process with $m_0=1$
for the Ising model and the Potts model
have been investigated with Monte Carlo simulations
\cite {sta92,mue93,li95a,sch95a}. 
The results show that universality and scaling appear to be
valid also in the early stage of the time evolution. 

In this letter we are more ambitious. We 
study whether there exists universal scaling behaviour 
in the short-time regime of the critical relaxation from
an initial state with very high temperature and 
{\it arbitrary magnetization}.
If the large time correlation length plays
an essential role for the universality in
 the critical dynamics
 as it was pointed out above,
the presentation of the universal behaviour should not rely on
from what initial conditions the critical relaxation starts, 
and only the scaling behaviour of the initial conditions should 
be considered very carefully. 
Since the arbitrarily valued initial magnetization
is no more around a fixed point, 
a critical exponent $x_0$ is not sufficient
to describe the scaling behaviour of 
the initial magnetization. In the language of 
the renormalzation group method, the effective 
dimension of the initial magnetization 
will in general  depend of the initial magnetization
itself. In order to describe this phenomena, we introduce
a {\it critical characteristic function} rather than
a critical exponent.
For the $k$-th moment of the magnetization,
 the generalized scaling relation may be written as
\begin{equation}
M^{(k)}(t,\tau,L,m_{0})=b^{-k\beta/\nu}
M^{(k)}(b^{-z}t,b^{1/\nu}\tau,b^{-1}L,
\chi (b,m_{0}))
\label{e15}
\end{equation}
For  convenience of later discussions, finite 
systems have been considered here and $L$ is the lattice size.
The scaling behaviour of the initial magnetization $m_0$
is specified by the critical characteristic function
$\chi (b,m_{0})$, which in the limit
$m_0 \rightarrow 0$ tends to the simple
form $b^{x_{0}}m_{0}$, but is in general different.
Such a generalized scaling form  is in a similar spirit
as that in the correction to the scaling,
where non-linear effects of an off-fixed point
are considered. Here only our initial magnetization
is a relevant operator
and can be far away from the fixed point.
The ansatz that the exponents $\beta$, $\nu$ and $z$
do not depend on the initial magnetization $m_0$
is based on the assumption that the initial conditions should
not enter the renormalization of the critical system
in equilibrium and near equlibrium.
In other words, if there is a scaling form in the
short-time regime of the dynamic relaxation process, 
the scaling form should smoothly cross over to that
in the long-time regime. This greatly simplifies
the short-time behaviour of the critical dynamics.
In the neighbourhood of $m_0=1$ the critical characteristic function 
$\chi (b,m_{0})$ may also be characterized by a critical exponent.
 One can also realize that 
in the limit of $b=0$, $\chi (b,m_{0}) \rightarrow 0$, 
and when $b$ approaches infinity, $\chi (b,m_{0}) \rightarrow 1$.
In this letter we are interested in the more
general case, i.e. $m_0$ is between $0$ and $1$, and
$b$ is a reasonable finite number. The limiting cases will be discussed
in detail elsewhere.

Here we stress that the scaling relation 
in (\ref{e15}) is not trivial even though the critical 
characteristic function $\chi (b,m_{0})$ looks not so 
{\it "simple"} as a critical exponent. The scaling relation  relates
the time evolution of the observables 
with different initial magnetizations to each other and
represents the {\it self-similarity} of the dynamic systems.
{\it All the physical observables as functions of
 $t$, $\tau$ and $L$ should be described by the same
critical characteristic function $\chi (b,m_{0})$}.
Besides this, the physical observables are
universal functions of the variables $t$, $\tau$ and $L$
up to a non-universal scaling constant 
and the dependence of the observables as well as 
$\chi (b,m_{0})$ on $m_0$ is expected to be universal
up to a rescaling of the initial magnetization $m_0$.

As a concrete example, we consider the two dimensional
 Potts model, for which a quite accurate
dynamic exponent $z$ has been obtained from 
the universal behaviour of the
short-time dynamics \cite {sch95,oka95}. 
The Hamiltonian for the $q$ state Potts model is given by
\begin{equation}
H=K  \sum_{<ij>}  \delta_{\sigma_i,\sigma_j},\qquad \sigma_i=1,...,q
\label{hami}
\end{equation}
where $<ij>$ represents nearest neighbors. In our notation 
the inverse temperature is already absorbed into the coupling $K$.
It is known that the critical
points locate at $K_c=log(1+\sqrt{q})$. In this paper 
the three state ($q=3$) case  will be investigated.
In generating the random initial configurations,
we have sharply prepared the initial magnetization 
in order to avoid extra finite size effects from
the fluctuation of $m_0$ in the finite systems.
After the preparation of the initial configurations
the system is released to evolve according to the
Heat-bath algorithm .

\begin{table}[h]\centering
$$
\begin{array}{|l|c|l|l|l|l|}
\hline
     &  m_0  &  \multicolumn{2}{c|} {0.14}   & \multicolumn{2}{c|} {0.40}  \\
\hline
 (L_1,L_2)  & &\quad M_d &\quad U_d &\quad M_d &\quad U_d \\
\hline
 (36,72) & \chi (2,m_0) &  0.1807(05) & 0.1804(08) & 0.5298(06)& 0.5300(36)\\
    (72,144) &  &  0.1800(04) & 0.1798(04) & 0.5307(06)& 0.5302(45)\\
\hline
(36,144) & \chi (4,m_0) & 0.2328(09) & 0.2324(11) & 0.6918(28) &  0.6910(78)\\
\hline
\end{array}
$$
\caption{ 
 $\chi (b,m_0)$ measured 
from the magnetization and the Binder-type cumulant.
}
\label{T1}
\end{table}

For simplicity, we take $\tau=0$ and
therefore the exponent $\nu$ will not enter the calculation.
The exact value of the static exponents
$\beta/\nu=2/15$ and the dynamic exponent $z=2.196$ 
obtained
from the power law decay of the auto-correlations
\cite {sch95,oka95} 
will be taken as input.
To verify the scaling relation (\ref{e15}) and
determine the critical characteristic function
$\chi (b,m_{0})$, we perform the simulation
for a pair of lattice sizes $L_1$
and $L_2$ and measure the time
evolution of the magnetization and the second moment defined as
\begin{equation}
M^{(k)}(t,m_0)= <\lbrack \frac{3}{2 N}\,\sum_i
      (\delta_{\sigma_i(t),1}-\frac{1}{3}) \rbrack ^{k}>, \ \ \ \ k=1,2
\end{equation}
where the average is taken over independent
initial configurations and the random forces.
In order to reduce extra errors from $M^{(k)}(t,1)$,
especially when $m_0$ becomes bigger, we introduce
a magnetization difference
\begin{equation}
M_d(t,m_0)= M^{(1)}(t,1)-M^{(1)}(t,m_0)
\end{equation}
and a Binder-type cumulant 
\begin{equation}
U_d(t,m_0)= \frac{M^{(2)}(t,1)-M^{(2)}(t,m_0)}{[M_d(t,m_0)]^2}.     
\end{equation}
From the scaling collaps of $M_d(t,m_0)$ or $U_d(t,m_0)$
 for two lattices
with suitable initial magnetizations,
we can estimate the values of the function $\chi (b,m_{0})$
at $b=L_2/L_1$ for different $m_0$.
 
\begin{figure}[t]\centering
\epsfysize=12cm
\epsfclipoff
\fboxsep=0pt
\setlength{\unitlength}{1cm}
\begin{picture}(13.6,12)(0,0)
\put(1.2,8.0){\makebox(0,0){$M_d(t)$}}
\put(11.8,1.2){\makebox(0,0){$t$}}
\put(8.4,7.){\makebox(0,0){\footnotesize$L=144, m_0=0.14$}}
\put(10.,4.8){\makebox(0,0){\footnotesize$L=72, m_0=0.175$}}
\put(9.2,3.1){\makebox(0,0){\footnotesize$L=72, m_0=0.185$}}
\put(0,0){{\epsffile{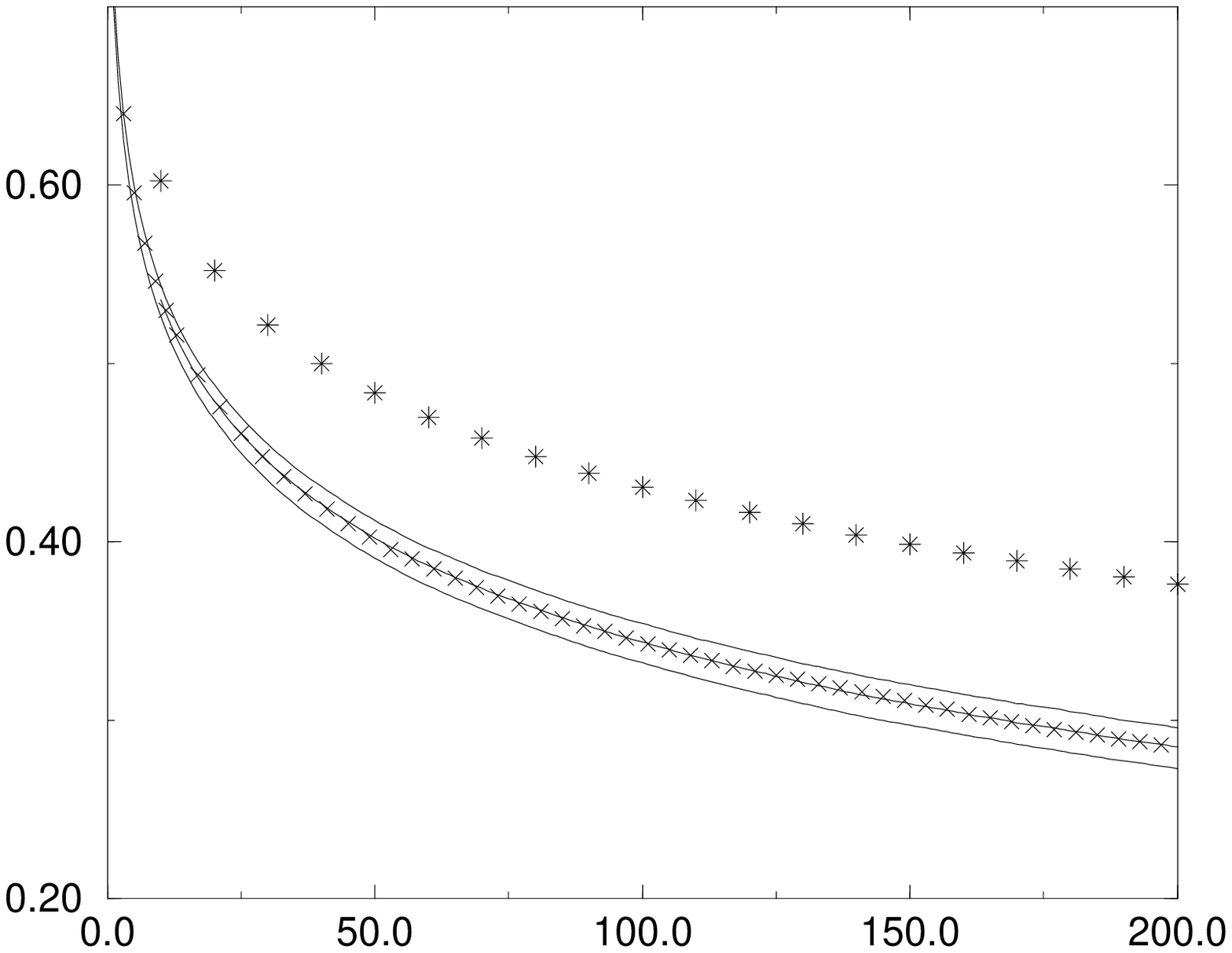}}}
\end{picture}
\caption{ The scaling plot for the magnetization with $(L_1,L_2)=(72,144)$.
}
\label{F1}
\end{figure}

In Fig. 1 the scaling plot for the magnetization
is displayed for the lattice $L_2=144$
with initial magnetization $m_{02}=0.14$
and the lattice $L_1=72$ with  suitably 
correponding initial magnetizations $m_{01}$.
The stars represent the time evolution of the magnetization
of the lattice $L_2=144$.
The crosses are the same data but rescaled in time and
multiplied by an overall factor $2^{\beta/\nu}$.
If one can find a $m_{01}$ for lattice
$L_1=72$ such that its time evolution of the magnetization
fits to the crosses, the scaling is valid and from
the scaling relation (\ref {e15}) one gets
$\chi (2,m_{02})=m_{01}$.
Practically we have performed the simulation for $L_1=72$
with two different initial magnetizations $m_0=0.175$
and $m_0=0.185$, for which the magnetizations
have been plotted by the lower and upper solid
 lines in Fig. 1. By linear extrapolation, we obtain
the time evolution of the magnetization with initial values
between these two values. Then we can estimate
$m_{01}$. 
The solid line laying on the crosses in Fig. 1 is the time evolution of
the magnetization for $L=72$
with an initial magnetization $m_0=0.1800(4)$ which
has the best fit to the crosses, i.e. $\chi (2,0.14)=0.1800(4)$.
From the previous numerical simulations for the short-time
dynamics \cite {li94,sch95,oka95}, we know that
the microscopic time scale $t_{mic}$ is negligibly small
for the Heat-bath algorithm, at least for the measurement
of the dynamic exponent $\theta$ or the exponent $x_0$. 
In our calculation we have carried out the
fitting procedure in a time interval of [10,200]
in the time scale of lattice $L=72$.
From Fig. 1 one can also see clearly that the scaling
relation is valid starting from the very early stage of the time
evolution. 

\begin{figure}[t]\centering
\epsfysize=12cm
\epsfclipoff
\fboxsep=0pt
\setlength{\unitlength}{1cm}
\begin{picture}(13.6,12)(0,0)
\put(1.2,8.4){\makebox(0,0){$M_d(t)$}}
\put(11.8,1.2){\makebox(0,0){$t$}}
\put(7.0,6.5){\makebox(0,0){\footnotesize$L=144, m_0=0.14$}}
\put(9.0,5.0){\makebox(0,0){\footnotesize$L=36, m_0=0.225$}}
\put(8.0,2.8){\makebox(0,0){\footnotesize$L=36, m_0=0.240$}}
\put(0,0){{\epsffile{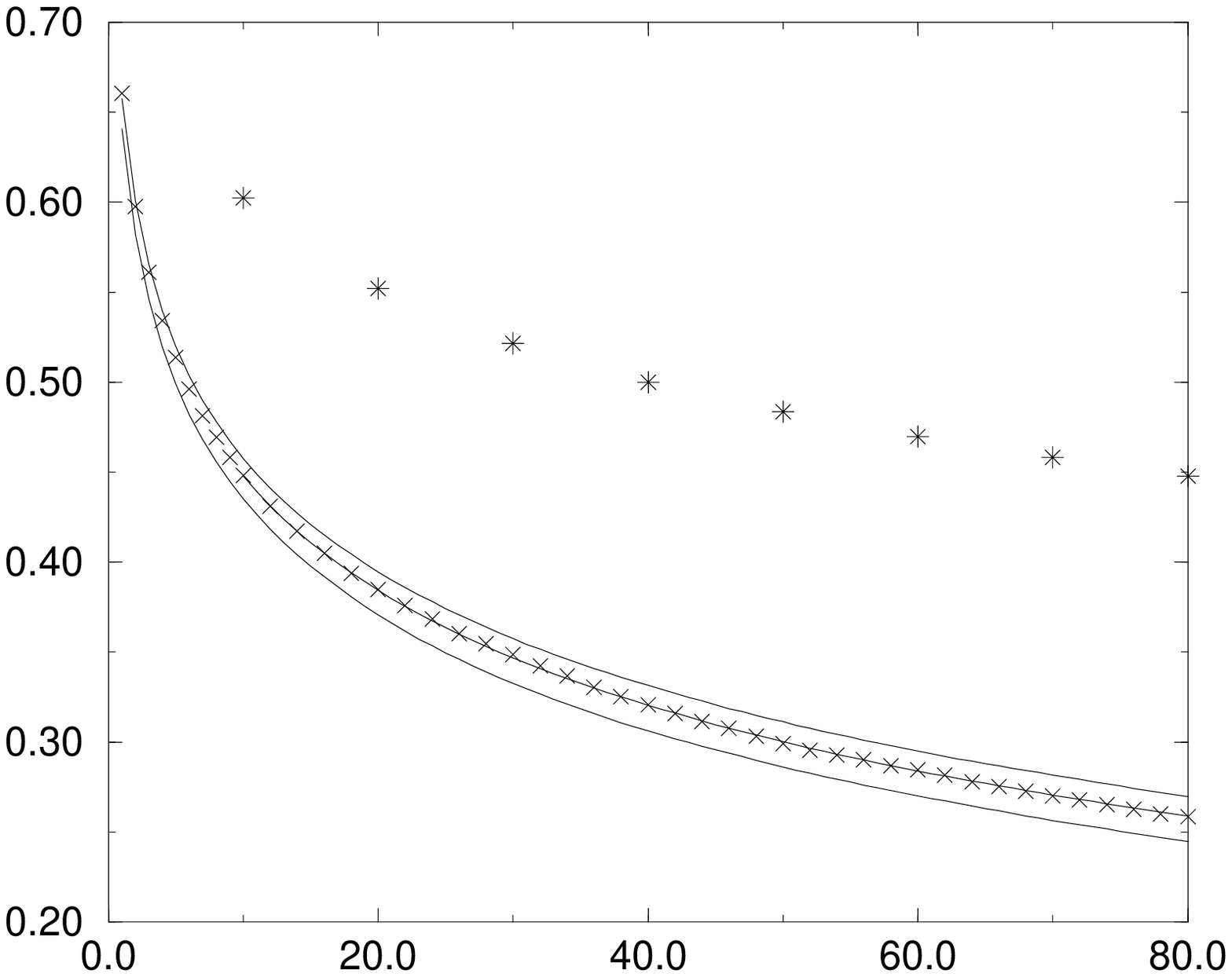}}}
\end{picture}
\caption{The scaling plot for the magnetization with $(L_1,L_2)=(36,144)$.
}
\label{F2}
\end{figure}

In Fig. 2 a scaling plot is shown
for the lattice $L_2=144$
with $m_{02}=0.14$ and $L_1=36$.
From a fitting in a time interval of [10,80] 
we obtained $\chi (4,0.14)=0.2328(9)$. 
In a similar way,  $\chi (b,m_0)$ may independetly obtained from
the scaling collaps of the Binder-type cumulant $U_d(t,m_0)$.
Altogether we have performed the simulation for
$m_{02}=0.14$, $0.22$, $0.30$, $0.40$, $0.50$, $0.60$, $0.70$,
$0.80$ and $1.00$.  
In Tab. 1 the results for two typical values
$m_{02}=0.14$ and $0.40$ are given. 
We can see that $\chi (b,m_0)$ estimated from both
$M_d(t,m_0)$ and $U_d(t,m_0)$
are very well consistent.
In order to see the finite size effect, we have also performed
the calculation for other pairs of lattices
as e.g. $(36,72)$. In Tab. 1
we can see that the finite size effect for the lattice
pair $(L_1,L_2)=(36,72)$ is already quite small.
In the simulations, the statistics of $L=36$, $L=72$ and
$L=144$ are respectively $80\ 000$, $40\ 000$ and $8\ 000$.
In Tab. 1, errors are estimated
by dividing the data into four groups.
 In principle the measurements
from the magnetization are more reliable than those from the
 Binder-type cumulants where higher moments
are involved, at least when the static exponent 
$\beta/\nu$ is known. The magnetization is
self-averaging, but the Binder-type cumulant is not.
When $m_0$ is getting bigger, $U_d(t,m_0)$ is more fluctuating.

In order to get a more direct understanding
 of the full critical characteristic function
$\chi (b,m_{0})$, we define
an effective dimension $x(b,m_0)$ of the initial magnetization
$m_0$ by $\chi (b,m_{0}) = b\ ^{x(b,m_0)}\ m_0$. 
By the definition naturally $x(b,0)=x_0$. From the values of
$\chi (b,m_{0})$,
we can calculate $x(b,m_0)$.
Taking the results from the magnetization difference with the lattice pair
$(L_1,L_2)=(36,144)$ and $(72,144)$, the corresponding 
effective dimensions $x(b,m_0)$ are plotted in Fig. 3.
It clearly shows that $\chi (b,m_{0})$ is a non-trivial function.
For example, the values $x(2,0.14)=0.363(3)$ and 
$x(2,0.40)=0.408(2)$
apparently differ from $x_0=0.298(6)$ \cite {oka95}.
When $m_0$ varies from zero to one, the
effective dimension $x(b,m_0)$ first increases
and then decreases to zero. However, when $m_0$ (or $\chi (b,m_0)$) is
approaching $m_0$(or $\chi (b,m_0)$)=1 it is not the best choice
to determine $x(b,m_0)$ directly in the way discussed 
in this letter since the dependence of the physical
observables on $m_0$ becomes weaker and
we face big statistical fluctuations.

\begin{figure}[t]\centering
\epsfysize=12cm
\epsfclipoff
\fboxsep=0pt
\setlength{\unitlength}{1cm}
\begin{picture}(13.6,12)(0,0)
\put(1.2,8.5){\makebox(0,0){$x(b,m_0)$}}
\put(12.2,1.2){\makebox(0,0){$m_0$}}
\put(11.,8.){\makebox(0,0){\footnotesize$b=2$}}
\put(9.3,5.2){\makebox(0,0){\footnotesize$b=4$}}
\put(0,0){{\epsffile{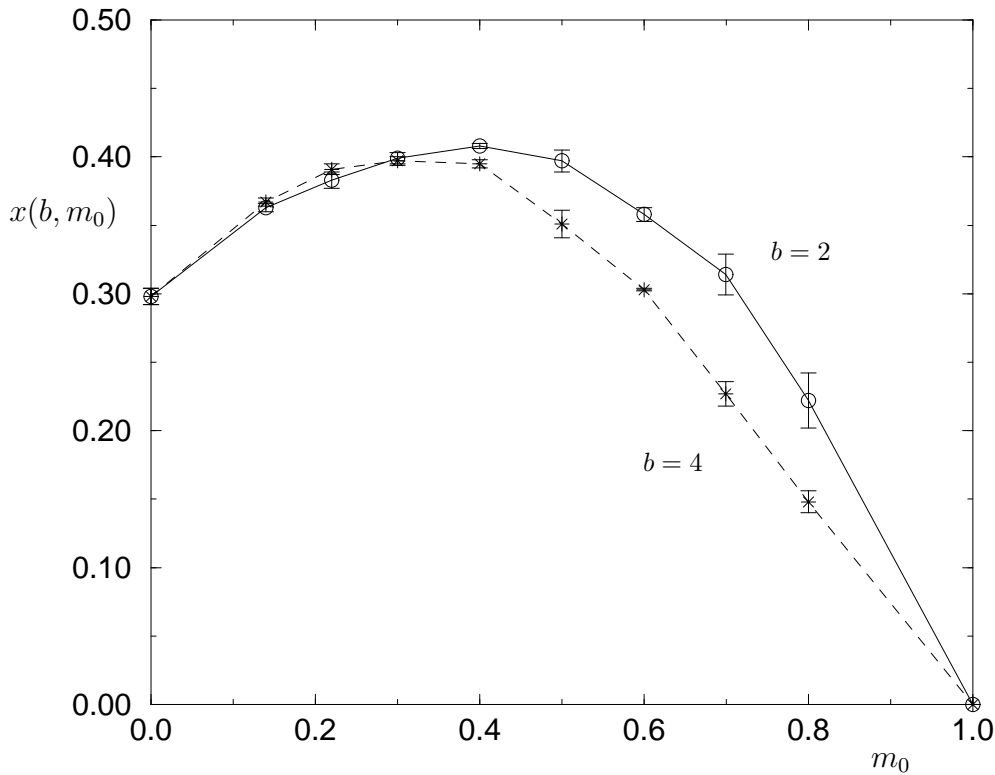}}}
\end{picture}
\caption{ The effective dimension obtained from the magnetization 
with $(L_1,L_2)=(72,144)$ and $(36,144)$. Circles are of $x(2,m_0)$
and stars are of $x(4,m_0)$. The lines are drawn to guide the eyes.
}
\label{F3}
\end{figure}

Finally let us have some more understanding of the
 dynamic system discussed above. 
Supposing $m_0$ is also a `coupling' of the system,
in the `$K-m_0$' plane of the couplings there exists a critical
line $K=K_c$ in the sense that the time correlation is
divergent. In the neighbourhood of this
critical line,
the time correlation length depends only on
the coupling $K$, and its scaling behaviour 
is characterized by the exponent $\nu z$. If the critical line is
a line of the fixed points, the exponent $\nu$ together with $z$ 
and $\beta$ is sufficient to describe the critical
dynamic system.
However, in our case the critical line is {\it not a line of the fixed
points}. Therefore a critical characteristic function
should in general  be introduced to describe the scaling behaviour.
It is actually interesting to see what happens
for a system in the equilibrium where a critical line exists
but the line is not a line of fixed points.

In conclusion, we have numerically simulated the universal
short-time behaviour
of the dynamic relaxation process for the two dimensional
critical Potts model starting from an initial state 
with very high temperature and arbitrary magnetization.
The results show that the traditional scaling relation should
be generalized. A critical characteristic function is
introduced to describe the scaling behaviour of the initial 
magnetization. We demonstrate how to determine
numerically the critical characteristic function. 
The study of the short-time
dynamics is not only conceptually interesting but also
pratically important since it is possible 
to obtain the static exponents $\beta$, $\nu$ and
the dynamic exponent $z$ of the critical systems
far before the dynamic process reaches the equilibrium.
It is challenging to derive analytically the generalized scaling
relation in (\ref{e15}) as well as the critical characteristic
function $\chi (b,m_{0})$.
The application to the {\it dynamic} field theory, e.g. the stochastic
quantization of the field theory, is attractive since
such a knowledge would be important for the numerical simulation
of the lattice gauge theory.

{\it Acknowledgement:} The author would like to thank
   L. Sch\"ulke, S. Marculescu and K. Okano for very helpful discussions and 
   K. Untch for maintaining the workstations.

\end{document}